\documentclass
[floatfix,superscriptaddress,secnumarabic,amssymb,amsmath,nobibnotes,aps,prd,showkeys,showpacs,nofootinbib,onecolumn,notitlepage,12pt]{revtex4}%
\usepackage{setspace}
\usepackage{color}
\usepackage{amsmath}
\usepackage{amsfonts}
\usepackage{verbatim}
\usepackage{amssymb}
\usepackage{graphicx,bm}
\usepackage{graphicx}
\usepackage{amsmath}
\usepackage{amssymb}
\usepackage{amssymb}
\usepackage{graphicx,bm}
\usepackage{graphicx}
\usepackage[caption=false]{subfig}%
\providecommand{\U}[1]{\protect\rule{.1in}{.1in}}
\newcommand{\be}{\begin{equation}}
\newcommand{\ee}{\end{equation}}

\newcommand{\mincir}{\raise
-3.truept\hbox{\rlap{\hbox{$\sim$}}\raise4.truept\hbox{$<$}\ }}
\newcommand{\magcir}{\raise
-3.truept\hbox{\rlap{\hbox{$\sim$}}\raise4.truept\hbox{$>$}\ }}

\ifx\pdfoutput\relax\let\pdfoutput=\undefined\fi
\newcount\msipdfoutput
\ifx\pdfoutput\undefined\else
\ifcase\pdfoutput\else
\ifx\paperwidth\undefined\else
\ifdim\paperheight=0pt\relax\else\pdfpageheight\paperheight\fi
\ifdim\paperwidth=0pt\relax\else\pdfpagewidth\paperwidth\fi
\fi\fi\fi
\begin{document}
\title{Cosmological Constant from Equivalent Transformation in Quantum Cosmology}
\author{Andronikos Paliathanasis}
\email{anpaliat@phys.uoa.gr}
\affiliation{Institute of Systems Science, Durban University of Technology, Durban 4000,
South Africa}
\affiliation{Departamento de Matem\'{a}ticas, Universidad Cat\'{o}lica del Norte, Avda.
Angamos 0610, Casilla 1280 Antofagasta, Chile}

\begin{abstract}
We explore the introduction of the cosmological constant via equivalent
transformations in cosmology. We consider the Wheeler-DeWitt equation for the
CDM universe and we construct the Hamilton-Jacobi action for the $\Lambda$CDM
model. We discuss how this approach allows us to relate different physical
systems, providing insights into the role of the cosmological constant in cosmology.

\end{abstract}
\maketitle

\section{\qquad Introduction}

\label{sec1}

The cosmological constant, $\Lambda$, was introduced by Einstein to enable the
static FLRW universe with cold dark matter (CDM), as provided by the theory of
General Relativity \cite{ein1}. Nowadays, the cosmological constant plays an
important role in describing the late-time acceleration phase of our universe
\cite{rr1,Teg,Kowal,Komatsu}. The $\Lambda$CDM is the simplest cosmological
model capable of addressing cosmological observations. Despite recent data
challenging the cosmological constant \cite{lper}, the $\Lambda$CDM model is
used as a reference for dark energy models.

The cosmological constant is associated with a constant dark energy density
(vacuum energy) \cite{ss1}; this density remains constant even in the case of
a nonstatic universe. However, this property gives rise to the "old
cosmological problem" \cite{Weinberg89}, where the quantum vacuum is estimated
to be more than a hundred orders of magnitude larger than the measured value
of the cosmological constant. On the other hand, today the vacuum density is
similar to the matter density, which is known as the coincidence problem
\cite{Pad03,coincidence}. For more details, we refer the reader to
\cite{cc1,cc2,cc3}. The $\Lambda$CDM universe describes two main eras: the CDM
matter epoch, where the universe is dominated by the dust fluid, and the de
Sitter epoch, which is a future attractor. The gravitational field equations
are second-order and can be linearized by a simple change of variables, making
the derivation of the analytic solution a straightforward task.

In this work we deal with the Wheeler-DeWitt (WDW) formulation of quantum
cosmology. The WDW equation follows from a canonical quantization of gravity;
it was introduced by De Witt \cite{deWitt} and Wheeler \cite{whe}. For
gravitational theories described by minisuperspace, the WDW equation is a
single differential equation also known as the Schr\"{o}dinger equation of
quantum cosmology, and it has been extensively studied previously, as seen in
\cite{ha1,ha2,qq1,qq2}. The WDW equation is not the only \textquotedblleft
quantum equation\textquotedblright\ of gravity, and its validation cannot be
rejected or confirmed by observations \cite{sww1}. Nevertheless, it has been
used in the past to study various issues in quantum cosmology \cite{sm1}, such
as the initial cosmological singularity \cite{qq8} and inflation \cite{win}.

We demonstrate the existence of a map, specifically a point transformation,
that links the WDW equations for the CDM and $\Lambda$CDM models. To
illustrate this, we employ an Eisenhart lift and express the field equations
for these two cosmological models as sets of geodesic equations in an extended
minisuperspace. By applying different quantum operators to solve the WDW
equation, we obtain wavefunctions corresponding to classical systems
representing the CDM and $\Lambda$CDM systems. This approach allows the
cosmological constant to emerge from quantum cosmology through the application
of a quantum operator. Consequently, we can construct new cosmological models
with varying values for the cosmological constant, starting from the $\Lambda
$CDM model. The structure of the paper is outlined as follows.

In Section \ref{sec2} we briefly discuss the Hamiltonian formalism for the
$\Lambda$CDM model and we introduce the Eisenhart-lift. The WDW equation is
introduced in Section \ref{sec3} and we show how the application of different
quantum operators lead to the different classical universes. Finally, in
Section \ref{sec6} we draw our conclusions.

\section{$\Lambda$CDM Cosmology}

\label{sec2}

We consider a homogeneous and isotropic universe described by the spatially
flat FLRW line element%
\begin{equation}
ds^{2}=-N^{2}\left(  t\right)  dt^{2}+a^{2}\left(  t\right)  \left(
dx^{2}+dy^{2}+dz^{2}\right)  . \label{dd.01}%
\end{equation}
Scalars $N\left(  t\right)  $, $a\left(  t\right)  $ represent the lapse
function and the scale factor, respectively. The volume of the
three-dimensional space is defined as $V\left(  t\right)  =a\left(  t\right)
^{3}$, while the Hubble function reads $H\left(  t\right)  =\frac{1}{N}%
\frac{\dot{a}}{a}$.

We consider the gravitational action to be that of General Relativity we a
cosmological constant term%
\begin{equation}
S_{GR}=\int dx^{4}\sqrt{-g}\left(  R-2\Lambda\right)  +\int dx^{4}\sqrt
{-g}L_{m} \label{let.02}%
\end{equation}
where $R$ is the Ricci scalar for the background geometry, $\Lambda$ is the
cosmological constant and $L_{m}$ is the Lagrangian for the cold dark matter.

For the FLRW spacetime (\ref{dd.01}) the gravitational field equations are%
\begin{equation}
3H^{2}-2\Lambda=\rho_{m}, \label{s.01}%
\end{equation}%
\begin{equation}
\frac{2}{N}\dot{H}+3H^{2}-\Lambda=0, \label{s.02}%
\end{equation}
while the conservation of energy for the cold dark matter reads%
\begin{equation}
\dot{\rho}_{m}+3\left(  \frac{\dot{a}}{a}\right)  \rho_{m}=0.
\end{equation}
The latter equation gives $\rho_{m}=\rho_{m0}a^{-3}$.

Equations (\ref{s.01}) and (\ref{s.02}) form a Lagrangian system with
point-like Lagrangian function%
\begin{equation}
L\left(  N,a,\dot{a}\right)  =\frac{3}{N}a\dot{a}^{2}+N\left(  2\Lambda
a^{3}+\rho_{m0}\right)  .
\end{equation}

Friedmann's equation (\ref{s.01}) is the Hamiltonian function. We define the
momentum $p_{a}=\frac{6}{N}a\dot{a}$, then it reads%
\begin{equation}
\mathcal{H}^{\Lambda}\equiv N\left(  \frac{p_{a}^{2}}{12a}-\left(  2\Lambda
a^{3}+\rho_{m0}\right)  \right)  =0. \label{let.03}%
\end{equation}

The Hamiltonian function is of the form $\mathcal{H}=N\left(  \gamma
^{AB}\left(  \mathbf{q}\right)  p_{A}p_{B}-V\left(  \mathbf{q}\right)
\right)  $, where $V\left(  \mathbf{q}\right)  $ is the potential function,
and $\gamma^{AB}\left(  \mathbf{q}\right)  $ is known as the minisuperspace.

\subsection{Eisenhart-lift}

The Eisenhart-lift describes a systematic method for the geometrization of
Hamiltonian dynamical systems \cite{en1}. In Eisenhart's geometric approach,
new additional degrees of freedom are introduced, which increase the dimension
of the space where the motion occurs, along with the simultaneous introduction
of conservation laws related to the isometries of the new extended geometry.
For some previous studies of the Eisenhart-lift in cosmology, see, for
instance \cite{en2,en3,en4}.

We introduce the Hamiltonian function%
\begin{equation}
\mathcal{H}_{lift}^{\Lambda}\equiv N\left(  \frac{p_{a}^{2}}{12a}+p_{u}%
p_{v}-\left(  \omega a^{3}+\mu\right)  p_{v}^{2}\right)  =0.
\end{equation}

The equations of motion are%
\begin{equation}
\frac{1}{N}\dot{a}=\frac{p_{a}}{6a}~,~\frac{1}{N}\dot{u}=p_{v}-2p_{u}%
~,~\frac{1}{N}\dot{v}=p_{u},
\end{equation}%
\[
\frac{1}{N}\dot{p}_{a}=3\omega a^{2}p_{u}~,~\frac{1}{N}\dot{p}_{u}%
=0~,~\frac{1}{N}\dot{p}_{v}=0,
\]
in which $p_{u}=p_{u}^{0}$ and $p_{v}=p_{v}^{0}$ are conservation laws.

The latter dynamical system describes the (null-) geodesic equations for the
three-dimensional minisuperspace with line element%
\begin{equation}
ds_{lift}^{2}=6ada^{2}+2dudv+2\left(  \omega a^{3}+\mu\right)  du^{2}.
\end{equation}
We remark that null-geodesics are invariant under conformal transformations,
thus the lapse function $N$ play no role on the dynamics. 

Moreover, application of the two conservation laws $p_{u}=p_{u}^{0}$ and
$p_{v}=p_{v}^{0}$ in $\mathcal{H}_{lift}^{\Lambda}$ leads to the the
Hamiltonian function $\mathcal{H}^{\Lambda}~$for the $\Lambda$CDM with
constraint equations~%
\begin{equation}
\rho_{m0}=\mu p_{v}^{0}-p_{u}^{0}p_{v}^{0}\text{ and }\Lambda=\frac{\omega}%
{2}\left(  p_{u}^{0}\right)  ^{2}.
\end{equation}

\subsection{Wheeler-De Witt equation}

When a minisuperspace description exist, the WDW equations follows from the
canonical quantization $p=-i\hbar\frac{\partial}{\partial q}$ of the
Hamiltonian function. Because the field equations are invariant under
conformal transformations meaning the definition of $N$ plays no role in the
dynamics and the conformal Laplace operator is considered instead of the
Laplace operator. Thus, the WDW equation belongs to the family of Yamabe equations.

Consider the Hamiltonian%
\begin{equation}
\mathcal{H}\equiv N\left(  \frac{1}{2}\gamma^{AB}\left(  q\right)  p_{A}%
p_{B}+V\left(  q\right)  \right)  =0 \label{hm.01}%
\end{equation}
where $\gamma^{AB}\left(  q\right)  $ is the minisuperspace of dimension
$n=\dim\gamma^{AB}$ and Ricci scalar $R_{\gamma}$ the WDW equation reads%
\begin{equation}
\Delta\Psi+\left(  \frac{n-2}{4\left(  n-1\right)  }R_{\gamma}\left(
q\right)  -2V\left(  q\right)  \right)  \Psi=0. \label{let.04}%
\end{equation}
Operator $\Delta$ is the Laplace operator defined as $\Delta=\frac{1}%
{\sqrt{\left\vert \gamma\right\vert }}\frac{\partial}{\partial q^{A}}\left(
\sqrt{\left\vert \gamma\right\vert }\gamma^{AB}\frac{\partial}{\partial x^{B}%
}\right)  $. For $n=2$, equation (\ref{let.04}) reduces to the usual
Klein-Gordon equation.

Let $\Psi\left(  q\right)  =A\left(  q\right)  e^{\frac{i}{\hbar}S\left(
q\right)  }$ be a solution of equation (\ref{let.04}), then in the WKB
approximation we end
\begin{equation}
\frac{1}{2}\gamma^{AB}\frac{\partial S}{\partial q^{A}}\frac{\partial
S}{\partial q^{B}}+V\left(  q\right)  -\frac{n-2}{8\left(  n-1\right)
}R_{\gamma}\left(  q\right)  =0,
\end{equation}
which is the Hamilton-Jacobi equation for the Hamiltonian function
(\ref{hm.01}) with the quantum correction which follows from the
minisuperspace scalar. In cosmological studies, the minisuperspace solves the
Yamabe problem, specifically the minisuperspace is conformally flat
\cite{ccd1}; which means that the quantum corrections related to $R_{\gamma}$
is eliminated.

\section{Introducing $\Lambda$ from quantum cosmology}

\label{sec3}

Consider now in the Eisenhart-lift the Hamiltonian for the CDM model, that
is,
\begin{equation}
\mathcal{H}_{lift}^{CDM}\equiv N\left(  \frac{p_{a}^{2}}{12a}+p_{u}p_{v}-\mu
p_{v}^{2}\right)  =0.\label{sd.01}%
\end{equation}
The corresponding WDW equation is
\begin{equation}
\frac{1}{6a}\left(  \frac{\partial^{2}}{\partial a^{2}}-\frac{1}{2a}%
\frac{\partial}{\partial a}\right)  \Psi+\frac{\partial}{\partial v}\left(
\frac{\partial}{\partial u}+\mu\frac{\partial}{\partial v}\right)
\Psi=0.\label{wd1}%
\end{equation}

There is an direct relation of the conservation laws for the classical system
and the symmetries of the Yamabe equation \cite{as21}. Specifically, from the
two conservation laws $p_{u}=p_{u}^{0}$ and $p_{v}=p_{v}^{0}$ we can construct
the quantum operators
\begin{equation}
\left(  i\frac{\partial}{\partial u}+p_{u}^{0}\right)  \Psi=0~,~\left(
i\frac{\partial}{\partial v}+p_{v}^{0}\right)  \Psi=0 \label{sd.02}%
\end{equation}
which gives%
\begin{equation}
\Psi=\bar{\Psi}\left(  a\right)  e^{-i\left(  p_{u}^{0}u+p_{v}^{0}v\right)  },
\end{equation}
where now
\begin{equation}
\bar{\Psi}\left(  a\right)  =\bar{\Psi}_{1}\exp\left(  4\sqrt{\frac{\rho_{m0}%
}{3}}a^{\frac{3}{2}}\right)  +\bar{\Psi}_{2}\exp\left(  -4\sqrt{\frac
{\rho_{m0}}{3}}a^{\frac{3}{2}}\right)  ~,~\rho_{m0}=\mu\left(  p_{v}%
^{0}\right)  ^{2}-p_{u}^{0}p_{v}^{0}.
\end{equation}

The wavefunction is written in the form$~\Psi\simeq e^{-iS_{CDM}\left(
a,u,v\right)  }$, where function~$S_{CDM}\left(  a,u,v\right)  $
\begin{equation}
S_{CDM}\left(  a,u,v\right)  =\pm4\sqrt{\frac{\rho_{m0}}{3}}a^{\frac{3}{2}%
}+\left(  p_{u}^{0}u+p_{v}^{0}v\right)  ,
\end{equation}
is nothing else than the action for the CDM model. Function~$S_{CDM}\left(
a,u,v\right)  ~$solves the Hamilton-Jacobi equation for the classical system
(\ref{sd.01}).

The WDW equation (\ref{wd1}) admits additional symmetries, which can be used
to construct quantum operators different from that of (\ref{sd.02}). There are
ten independent nontrivial linear operators which can constructed by the
symmetries of equation (\ref{wd1}). These linear operators are related to the
conformal symmetries of the three-dimensional flat minisuperspace.

Consider the conformal symmetry of the minisuperspace
\begin{equation}
X=2ua\frac{\partial}{\partial a}+3\left(  1+u^{2}\right)  \frac{\partial
}{\partial u}+\left(  2\frac{\mu}{\beta}+3+6\mu-\left(  3\mu u^{2}%
+2a^{3}\right)  \right)  \frac{\partial}{\partial v}\label{sd.005}%
\end{equation}
where $\beta$ is a constant. 

We determine the canonical variables%
\begin{equation}
a=\frac{\alpha}{\cos\left(  \sqrt{\frac{3}{2}\beta}U\right)  ^{\frac{2}{3}}%
}~,~u=\frac{1}{\sqrt{\mu}}\tan\left(  \sqrt{\frac{3}{2}\beta}U\right)
~,~v=\sqrt{\frac{2\mu}{3\beta}}\left(  \mu U+V\right)  -\frac{\sqrt{\mu}}%
{3}\left(  3+4\alpha^{3}\right)  \tan\left(  \sqrt{\frac{3}{2}\beta}U\right)
.
\end{equation}
The WdW equation (\ref{wd1}) reads
\begin{equation}
\frac{1}{6\alpha}\left(  \frac{\partial^{2}}{\partial\alpha^{2}}-\frac
{1}{2\alpha}\frac{\partial}{\partial\alpha}\right)  \Psi+\frac{\partial
}{\partial V}\left(  \frac{\partial}{\partial U}+\left(  \mu-2\beta\alpha
^{3}\right)  \frac{\partial}{\partial V}\right)  \Psi=0, \label{hj.02}%
\end{equation}
and the conformal symmetry becomes $X=\partial_{V}$. Thus, from the symmetry
$\partial_{U}$ we can define the quantum operators%
\begin{equation}
\left(  i\frac{\partial}{\partial U}+p_{U}^{0}\right)  \Psi=0~,~\left(
i\frac{\partial}{\partial V}+p_{V}^{0}\right)  \Psi=0,
\end{equation}
that is, $\Psi\left(  \alpha,U,V\right)  =\hat{\Psi}\left(  \alpha\right)
e^{-i\left(  p_{U}^{0}U+p_{V}^{0}V\right)  }$.

In the WKB approximation,~$\hat{\Psi}\left(  \alpha\right)  \simeq
e^{iS_{\Lambda}\left(  \alpha\right)  }$, from (\ref{hj.02}) we find the
Hamilton-Jacobi equation%
\begin{equation}
-\frac{1}{12\alpha}\left(  \frac{\partial S_{\Lambda}}{\partial\alpha}\right)
^{2}+p_{U}^{0}p_{V}^{0}-\left(  \mu-\beta\alpha^{3}\right)  p_{V}^{2}=0,
\end{equation}
equivalently,%
\begin{equation}
\frac{1}{12\alpha}\left(  \frac{\partial S_{\Lambda}}{\partial\alpha}\right)
^{2}-\left(  2\Lambda\alpha^{3}+\rho_{m0}\right)  =0, \label{sd.05}%
\end{equation}
with constraints%
\begin{equation}
~2\Lambda=-\beta p_{V}^{0}~\text{and~}\rho_{m0}=\mu\left(  p_{v}^{0}\right)
^{2}-p_{u}^{0}p_{v}^{0}~.
\end{equation}

Equation (\ref{sd.05}) is nothing else than the Hamilton-Jacobi equation for
the $\Lambda$CDM model (\ref{let.03}) with scale factor $\alpha\left(
t\right)  $. 

In a similar way we can apply similar transformations and construct new
solutions for the WDW equations which describe universes with different value
of the cosmological constant. This is an important observation because it can
relate the solution space for universes with different values of the
cosmological constant. 

We have not mentioned that when $\mu\left(  p_{v}^{0}\right)  ^{2}-p_{u}%
^{0}p_{v}^{0}~=0$, that is, $\rho_{m0}=0$, then we can say that we can
introduce the cosmological constant from the vacuum, and to derive the de
Sitter solution from the vacuum space.

\section{Conclusions}

\label{sec6}

In Newtonian physics, small oscillations are described by a linear
second-order differential equation with solutions that are periodic functions.
According to Sophus Lie theory, there exists a point transformation where the
small oscillations are described by the free particle equation \cite{line}.
Indeed, with the use of nonlinear methods, small oscillations can be depicted
by a linear line in a time-position diagram. Although this is a mathematical
construct, it establishes a relation between the shared conservation laws of
the two classical systems.

This exact equivalence between the two different physical systems was applied
in this study to introduce the cosmological constant. We considered the CDM
universe, and we found that from the WDW equation, we can construct the
Hamilton-Jacobi action for the $\Lambda$CDM model. The assumption that we have
done is the field equations to be described by an extended minisuperspace as
given by the Eisenhart-lift and to apply extend the application of quantum
operators of the WDW equation. The relaxation of the first-class constraints
of quantum cosmology discussed recently in \cite{kk1}.

Equivalent transformations have been identified before in cosmology
\cite{bjar0,bjar}. It was found that certain families of inflationary models
are related through point transformations.

\textbf{Data Availability Statements:} Data sharing is not applicable to this
article as no datasets were generated or analyzed during the current study.

\begin{acknowledgments}
AP thanks the support of VRIDT through Resoluci\'{o}n VRIDT No. 096/2022 and
Resoluci\'{o}n VRIDT No. 098/2022 and of the National Research Foundation of
South Africa.
\end{acknowledgments}

\end{document}